# Simulation of conventional cold-formed steel sections formed from Advanced High Strength Steel (AHSS)

Hamid Foroughi[1] and Benjamin W. Schafer[2]

**Abstract**
The objective of this paper is to explore the potential impact of the use of advanced high strength steel (AHSS) to form traditional cold-formed steel structural members. In this study, shell finite element models are constructed, and geometric and material nonlinear collapse analysis performed, on simulated lipped channel cross-section cold-formed steel members roll-formed from AHSS. AHSS sheet is currently being used in automotive applications with thickness ranging from 0.35 to 0.8 mm (0.0138 to 0.0315 in.) and yield strengths from 350 to 1250 MPa (51 to 181 ksi). However, AHSS has not yet been employed in cold-formed steel construction. To assess the impact of the adoption of AHSS on cold-formed steel member strength a group of forty standard structural lipped channel cross-sections are chosen from the Steel Framing Industry Association product list and simulated with AHSS material properties. The stress-strain models used in this study are based on AHSS in production, including dual-phase and martensitic steels. The simulations consider compression with work on bending about the major axis in progress. Three different bracing conditions are employed so that the impact of local, distortional, and global buckling, including interactions can be explored. Due to the higher yield stresses of AHSS the potential for interaction and mode switching is anticipated to be greater in these members compared with conventional mild steels. The simulations provide a direct means to assess the increase in strength created by the application of AHSS, while also allowing for future exploration of the increase in buckling mode interaction, imperfection sensitivity, and strain demands inherent in the larger capacities. The work is intended to be an initial step in a longer-term effort to foster innovation in the application of new steels in cold-formed steel construction.

[1] Graduate Research Assistant, Dept. of Civil Engineering, Johns Hopkins University, hforoug1@jhu.edu

[2] Professor, Dept. of Civil Engineering, Johns Hopkins University, schafer@jhu.edu

# 1. Introduction

Cold-formed steel (CFS) structural members use cold bent sheet steel to provide efficient structural shapes that are noncombustible and highly structurally efficient. Use of cold-formed steel members continues to grow for both architectural and structural applications in building construction. Due to the manufacturing process thicknesses are naturally limited and ultimate strength of practical shapes is thus necessarily limited to a relatively modest value. Advanced High Strength Steels (AHSS) have been developed for the automotive industry over the last 20 years (Keeler and Kimchi 2014). AHSS sheet has yield stresses as high as 1250 MPa and are able to maintain large ultimate tensile elongations (>10%) even for these high yield stresses. As a result AHSS can be readily formed/manufactured and supply material yield values that are significantly in excess of current applications in CFS building construction. With this new strength comes new potential for design, particularly in mid-rise applications for CFS structural members.

Existing design specifications, such as the Direct Strength Method (DSM) in AISI S100-12, provide a potential design framework for CFS members formed from AHSS, but the validity of the provided rules has not been substantiated for higher strength steels. AISI S100-12 strength predictions include local-global (L-G) interaction, but based on experimental results at the time (Schafer 2002, Schafer 2008) excluded local-distortional (L-D), distortional-global (D-G), and local-distortional-global (L-D-G) interaction. More recent experimental research has shown that for higher strength (non AHSS) steels, such as G550 ($F_y$=550 MPa (80 ksi)), L-D interaction should be included (Yap and Hancock 2008, 2011). The interaction becomes more pronounced because of the additional local post-buckling demands, and because higher strength sections use thinner sheet steel requiring additional intermediate stiffeners, further complicating the response. Lead by Prof. Camotim at TU-Lisbon, Yap and Hancocks's findings motivated new activity in buckling mode interaction for members. They began by using shell FE simulations and demonstrated conditions where L-D interaction may be significant (Dinis et al. 2011, Camotim and Dinis 2011, Silvestre et al. 2012). They then collaborated on a small test series on G550 columns to physically demonstrate L-D interaction and its related strength erosion (Young et al. 2013). Finally, additional simulations on simple and complex cross-sections and proposed design recommendations were provided (Dinis et al. 2014, Dinis and Camotim 2015, Martins et al. 2015, 2016). Work remains though as the AHSS steel grades have unique material response not captured in the studies on low ductility high strength G550 steel – and agreed upon solutions for handling these interactions are still needed.

Shell finite element (FE) models have proven to be a robust tool for exploring the strength of thin-walled cold-formed steel members (Schafer et al. 2010, Foroughi et al. 2014). Therefore, they are employed here to examine the impact of AHSS material properties on CFS member strength. This brief paper first summarizes the numerical modeling conducted, including boundary conditions, element choice, material model, and imperfections. Next the paper provides a summary of the strength results in comparison with the AISI-S100 strength predictions employing DSM. This summary is followed by brief comments on the results and a discussion of future work. This study is an initial step towards a longer-term study on the application of AHSS in CFS buildings and the development of improved design guidance.

## 2. Numerical modeling
Shell finite element models of cold-formed steel members are developed and analyzed in ABAQUS. The based details of the developed models are provided in this section.

A total of six types of AHSS including both dual-phase and martensitic steels are considered here, as shown in Figure 1. In addition, a mild steel with yield stress 207 Mpa (30 Ksi) has also been selected for baseline comparison. To aid in comparisons across the materials the elastic modulus has been set to 203,500 MPa (29500 ksi) and the Poisson's ratio to 0.3 for all materials. Note, for application in ABAQUS the engineering stress-strain is converted to true stress-strain.

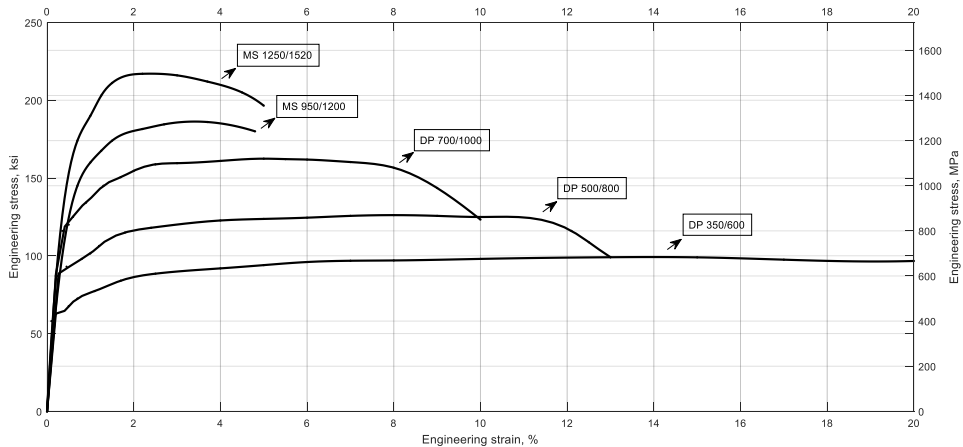

Figure 1. Engineering stress-strain curves for different type of AHSS (Keeler and Kimchi (2014))

Forty standard structural lipped channel cross-sections from the Steel Framing Industry Association (SFIA) product manual are selected for study as detailed in Table 1. The nomenclature, e.g. 250S162-33 provides the nominal wed depth in inches 250=2.5 in. (63.5mm), the flange width in inches 162=1.62 in. (41.1mm), and the material designation thickness in mils 33 mils = 0.033 in. (0.84 mm) – note designation thickness is not equal to design or minimum delivered thickness. Round corners (r = 2t per SFIA) are considered in the models. The members are meshed using the quadratic shell element S9R5 in ABAQUS with a mesh density consistent with Figure 2. The length of the studied members is three times the critical length for distortional buckling, as determined by CUFSM with simply-supported end boundary conditions.

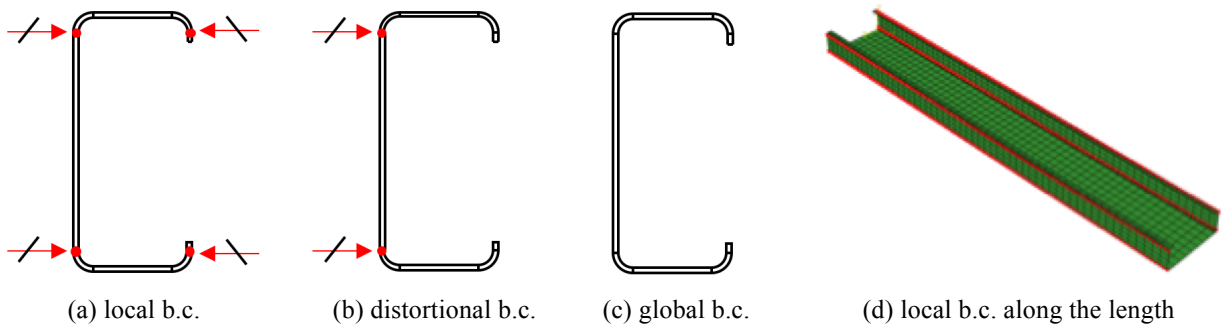

(a) local b.c.      (b) distortional b.c.      (c) global b.c.      (d) local b.c. along the length
Figure 2. Depiction of cross-section boundary conditions (b.c.) to establish different models

To help separate the buckling response of the members three cross-section boundary conditions are explored as depicted in Figure 2. The local model (Figure 2a) allows only local buckling in the cross-section. The distortional model (Figure 2b) allows local and/or distortional buckling. The global model (Figure 2c) allows local, and/or distortional, and/or global buckling. Thus, the local model isolates L; the distortional model allows L, D, and L-D; while the global model allows all possibilities. The end boundary conditions in the model are locally pinned. All translation degrees of freedom are tied to a reference node at the shear center. The reference node is pinned and restricted from torsion at one end and restricted from in-plane translation at the other end. All actions on the model are applied at the reference node.

Table 1. List of cold-formed steel cross sections for parametric study

| Designation | Thickness | | | | |
|---|---|---|---|---|---|
| | 33 | 43 | 54 | 68 | 97 |
| 250S162 | ✓ | ✓ | ✓ | ✓ | |
| 250S137 | | | ✓ | | |
| 350S162 | ✓ | ✓ | ✓ | | |
| 362S137 | | | | ✓ | |
| 362S162 | | | | ✓ | |
| 362S200 | | | ✓ | | |
| 400S137 | | | ✓ | | |
| 400S162 | | | | ✓ | |
| 400S200 | ✓ | ✓ | ✓ | | |
| 550S162 | ✓ | ✓ | ✓ | | |
| 600S137 | | ✓ | ✓ | ✓ | |
| 600S162 | | | | ✓ | ✓ |
| 600S200 | ✓ | ✓ | ✓ | | ✓ |
| 800S137 | | | ✓ | ✓ | |
| 800S162 | | | | ✓ | |
| 800S200 | ✓ | | | | |
| 800S250 | | ✓ | ✓ | | |
| 1000S162 | | ✓ | | | |
| 1000S200 | | | | | ✓ |
| 1200S162 | | | ✓ | | |
| 1200S200 | | | | | ✓ |
| 1200S250 | | | ✓ | | |
| 1200S162 | | | | ✓ | |

A surrogate finite strip model in CUFSM is used to generate geometric imperfections. A linear combination of the first local, distortional, and global buckling mode are used as the geometric imperfection distribution. The imperfection magnitude employs the 50% probability of exceedance values from Zeinoddini and Schafer (2012), as given in Table 2.

Table 2. Selected 50% imperfection exceedance values

| CFD | Local ($\delta/t$) | Distortional ($\delta/t$) | Global | | |
|---|---|---|---|---|---|
| | | | Bowl ($L/\delta$) | Camber ($L/\delta$) | Twist (Deg/m) |
| 50% | 0.31 | 0.75 | 2909 | 4010 | 0.30 |

Residual stresses and strains are not considered in the models at this time.

## 3. Simulation results and discussion

For each studied cross-section the member is compressed until a peak strength ($P_{FE}$) is achieved. Although complete response is of interest for future work, here only $P_{FE}$ is considered. For the local buckling boundary condition (Figure 2a), across the 40 studied members, and the 5 AHSS material grades and one mild steel grade, the normalized $P_{FE}$ as a function of local slenderness is provided in Figure 3 and compared with the DSM strength prediction for local buckling alone ($P_{nLo}$). $P_{nLo}$ is determined from the DSM expressions in AISI S100-12 with the global strength set equal to the squash load of the column ($P_y$). The results indicate good, but slightly conservative agreement for the DSM prediction in isolated local buckling. For isolated local buckling the models do not indicate a significant need for change in strength prediction to incorporate AHSS.

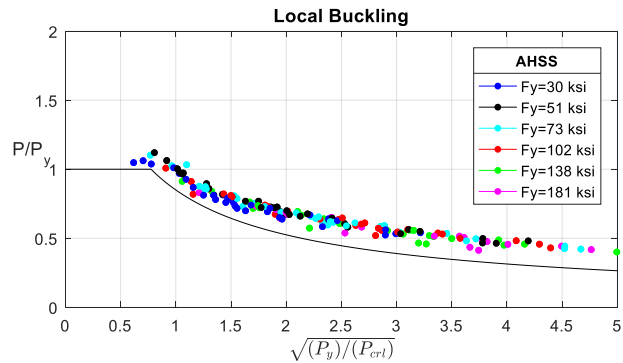

Figure 3. Comparison between DSM and AHSS member results under local boundary conditions of Figure 2a

For the distortional buckling boundary condition (Figure 2b) across the studied members the strength results are provided in Figure 4. If DSM predicts isolated local buckling controls ($P_{nLo}<P_{nD}$) then the $P_{FE}$ results are plotted against the DSM local curve, as in Figure 4a. If distortional buckling is predicted to occur, per existing DSM rules ($P_{nD}<P_{nLo}$) then the $P_{FE}$ are plotted against the distortional curve, as in Figure 4b. The results indicate that (1) distortional buckling is predicted to be far more prevalent in the AHSS materials with higher $F_y$ and (2) the DSM distortional curve is generally a good predictor of strength, but greater scatter exists in comparison with local buckling. The DSM design method as provided in AISI S100-12 and examined here does not consider L-D interaction. For the studied cross-sections, boundary conditions, member length, and material grades the interaction appears weak. (Note, $P_{crL}/P_{crD}$ ranges from 0.3 to 1.2, and 11 of the 40 studied members have $0.9<P_{crL}/P_{crD}<1.1$).

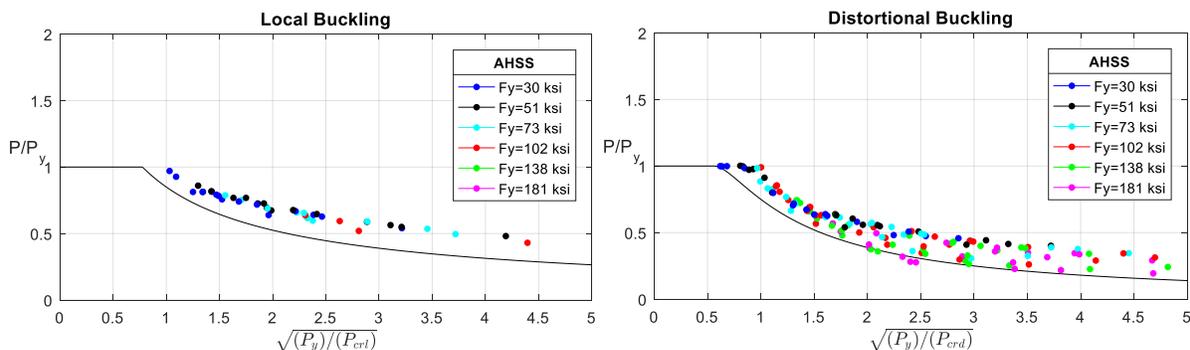

Figure 4. Comparison between DSM and AHSS member results under distortional boundary conditions of Figure 2b for specimens predicted as (a) local buckling controlled, and (b) distortional buckling controlled.

For the global buckling boundary condition (Figure 2b) across the studied members the strength results are provided in Figure 5. For this boundary condition L, D, G or any combination thereof can theoretically occur. DSM, per AISI S100-12 considers the strength as the minimum of L-G ($P_{nL}$) and D ($P_{nD}$). If DSM predicts L-G controls ($P_{nL}<P_{nD}$) then the $P_{FE}$ results are plotted against the DSM local curve, as in Figure 5a. If distortional buckling is predicted to occur, per existing DSM rules ($P_{nD}<P_{nL}$) then $P_{FE}$ are plotted against the distortional curve, as in Figure 5b. The results indicate (1) distortional buckling rarely is predicted to control for longer unbraced members, (2) significant scatter exists in the local-global strength prediction when visualized as a function of local slenderness alone.

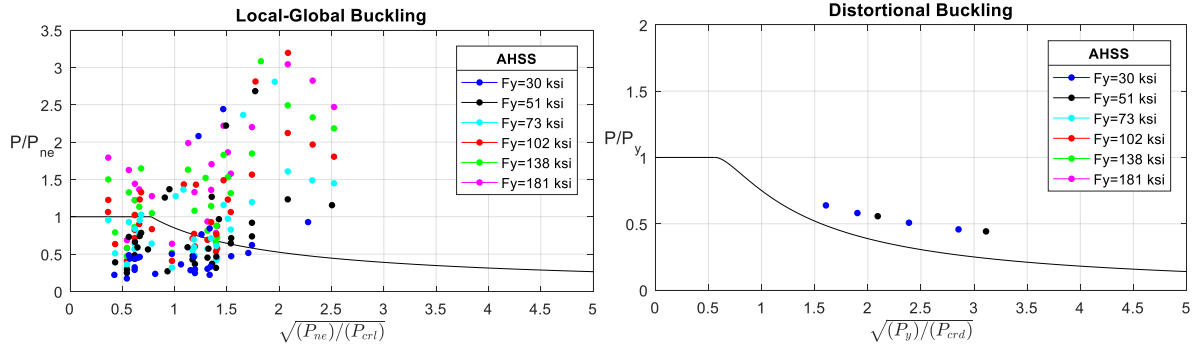

Figure 5. Comparison between DSM and AHSS member results under global "free" boundary conditions of Figure 2c for specimens predicted as (a) local-global buckling controlled, and (b) distortional buckling controlled.

To examine this second point regarding visualization of the predictions as a function of local slenderness, the $P_{FE}$ results that are predicted to be controlled by L-G interaction, are instead plotted as a function of global slenderness in Figure 6a. The results indicate the expected overall trend against global slenderness and the necessity to consider L-G interaction as the strength falls below the global strength prediction. In DSM L-G interaction is a function of the local slenderness ($\lambda_L=(P_{ne}/P_{crL})^{0.5}$) to see how this impacts the column strength the reduced global strength for different values of $\lambda_L$ are provided. The selected values of $\lambda_L$ include the mean +/- a standard deviation for the studied values of $\lambda_L$ as depicted in Figure 5a and in the histogram of $\lambda_L$ itself in Figure 6b. From this we can conclude that the large scatter observed in Figure 5a is somewhat misleading as it is normalized to $P_{ne}$, which is often a small capacity. The overall agreement with the L-G data is better shown against global slenderness as in Figure 6a. Nonetheless the long column results suggest that further refinement may be needed. In particular, D-G interaction is ignored, to what extent might this be influencing the scatter in the results? The results on columns which may fail in only L or D (Figure 4) indicate a large number of high yield stress AHSS sections fail in D – why would it be expected that essentially all of these would fail in L-G interaction for longer columns. Consistent with previous research findings of others, current DSM provisions may need to be revisited for D-G interaction, particularly for very high strength steel shapes, as is common with AHSS.

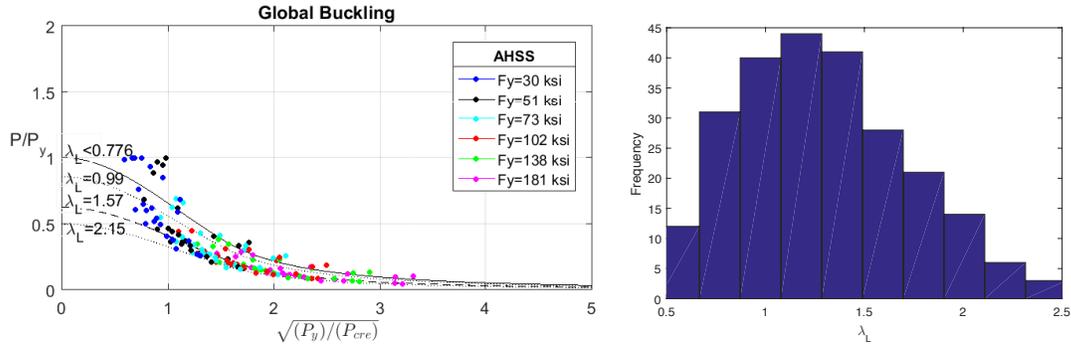
Figure 6. (a) Predicted strength as a function of global slenderness for Figure 5a specimens including reduced curves depending on local slenderness and (b) distribution of local slenderness in studied members

Significant work remains, even for the member-based FE collapse studies pursued herein. Pre-peak stiffness, post-peak response, local strain demands, impact of residual stresses and strains, impact of imperfection magnitude and distribution, examination of different loading actions including combined actions, detailed examination of existing and new options for considering interaction buckling in the strength prediction, reliability studies, and consideration of other cross-sections to name a few. Nonetheless, this initial study indicates that cold-formed steel members formed from AHSS do not radically depart from expected strengths, but additional studies may needed to fine tune existing design specifications when very high strength AHSS members are employed.

## 4. Conclusions

An initial examination of existing design strength predictions for cold-formed steel compression members formed from advanced high strength steel (AHSS) grades indicates that modest changes in the strength prediction methods will likely be needed if AHSS is adopted for steel construction. For members failing in isolated local or distortional buckling existing strength predictions from the Direct Strength Method in AISI S100-12 appear adequate. For longer members local-global interaction does not appear sufficient to accurately capture the strength and distortional-global interaction deserves further investigation. The finding is limited to the context of the study, which considers 5 AHSS grades, 40 commercially available cold-formed steel lipped channel cross-sections, at one member length, in compression, across a variety of cross-section boundary conditions. The results are arrived at through geometric and material nonlinear collapse shell finite element simulations performed in ABAQUS. Additional work is needed to explore additional loading actions, examine the strain demands inherent in the new sections, consider additional manufacturing impacts such as residual stresses and strains, and delve more deeply into buckling mode interaction issues for these members.

**Acknowledgements**
The authors would like to thank the American Iron and Steel Institute for partial support of this work. All findings and recommendations are those of the authors and do not necessarily reflect the views of the sponsor.